# PoolTestR: An R package for estimating prevalence and regression modelling with pooled samples

Angus McLure[1]*, Ben O'Neill[1], Helen Mayfield[1], Colleen Lau[1], Brady McPherson[1,2]

[1] Research School of Population Health, Australian National University, Canberra, Australia.
[2] Australian Defence Force Malaria and Infectious Disease Institute, Brisbane, Australia
* Corresponding author: angus.mclure@anu.edu.au
62 Mills Road,
Acton, ACT 2611,
Australia

**Software availability**

Name of software: PoolTestR
Type of software: Add-on package for R
First available: 2020
Programming languages: R, stan
License: GPL 3
Code Repository: CRAN (pending approval); https://github.com/AngusMcLure/PoolTestR
Developers: Angus McLure
Contact Address: angus.mclure@anu.edu.au

# Abstract

Pooled testing (also known as group testing), where diagnostic tests are performed on pooled samples, has broad applications in the surveillance of diseases in animals and humans. An increasingly common use case is molecular xenomonitoring (MX), where surveillance of vector-borne diseases is conducted by capturing and testing large numbers of vectors (e.g. mosquitoes). The R package PoolTestR was developed to meet the needs of increasingly large and complex molecular xenomonitoring surveys but can be applied to analyse any data involving pooled testing. PoolTestR includes simple and flexible tools to estimate prevalence and fit fixed- and mixed-effect generalised linear models for pooled data in frequentist and Bayesian frameworks. Mixed-effect models allow users to account for the hierarchical sampling designs that are often employed in surveys, including MX. We demonstrate the utility of PoolTestR by applying it to a large synthetic dataset that emulates a MX survey with a hierarchical sampling design.

# Highlights

- PoolTestR is an R package for analysing pooled testing (group testing) data
- PoolTestR can estimate prevalence and fit flexible mixed-effect regression models
- PoolTestR is flexible, extensible, and easy-to-use with small or large datasets
- PoolTestR is suited for hierarchical sampling designs, e.g molecular xenomonitoring

# 1 Introduction

Pooled testing, also known as group testing, where diagnostic tests are performed on pooled samples, has broad applications for the detection of traits, defects, or diseases with low prevalence. Recently, pooled testing has been used to efficiently and rapidly test for SARS-CoV-2019 (Sunjaya and Sunjaya, 2020), but this approach has long been used for surveillance of other infectious diseases, e.g. to detect pathogens in food animals (Arnold et al., 2011) and conduct surveillance of vector-borne diseases (Pilotte et al., 2017; Rodríguez-Pérez et al., 2006). The World Health Organization has for decades been running global elimination programs to reduce the impact of many vector-borne diseases, including lymphatic filariasis (LF) (World Health Organisation, 2019)(ref). A key challenge for programs of this scale is to optimise the efficiency and accuracy of surveillance, especially as the disease becomes rarer or more localised over time. Molecular xenomonitoring (MX), the surveillance of pathogens or molecular markers in vector populations, is emerging as an alternative or adjunct to human-based surveillance of vector-borne diseases such as LF, and typically employs pooled testing of mosquitoes for filarial DNA (Lau et al., 2016; Rao et al., 2016; Schmaedick et al., 2014; Subramanian et al., 2020). In the right setting, MX with pooled testing could potentially be very sensitive, avoid the need to test humans, and provide insights to assist in vector control strategies.

Large scale MX surveys, such as those needed to support decision making for elimination programs, involve capturing and testing large numbers of vectors. It is not always feasible to individually test every vector captured, so vectors are routinely pooled to reduce cost and improve efficiency. If using a sufficiently sensitive and specific test, each pool returns a positive result if any individual in the pool is positive and a negative result otherwise. The presence/absence of infected vectors provides a proxy measure of ongoing transmission, and could potentially help identify geographic areas where further surveillance or interventions may be required. The prevalence of infection amongst vectors provides useful information for decision makers when assessing and communicating risk, and can be used to prioritise the allocation of resources, observe the effects of interventions, and identify spatial and temporal trends. However, estimating prevalence from pooled samples requires specialised statistical methods (Chen and Swallow, 1990; Farrington, 1992; Hepworth, 2005; Walter et al., 1980)

Many MX surveys employ a hierarchical sampling structure. For instance, to conduct a MX study in a particular country, one may select a number of representative regions in a country, followed by a number of villages within each region, and place traps at a number of representative sites within each village. While hierarchical sampling designs like this provide an effective means of collecting a representative sample of vectors across the study area, they call for specialised analytical methods to accurately estimate infection prevalence. Analytical methods that do not account for hierarchical sampling structures will tend to underestimate the uncertainty in prevalence when applied to data collected using these sampling methods (Birkner et al., 2013).

There are a number of extant software packages for working with pooled testing models — e.g. PoolScreen (Katholi and Unnasch, 2006), and the `R` packages `pooling` (Van Domelen, 2020), `binGroup` (Zhang et al., 2018), and its successor `binGroup2` (Hitt et al., 2020). PoolScreen is a stand-alone application that has been used in many MX projects and provides a graphical user interface for estimating prevalence in both frequentists and Bayesian paradigms. However, PoolScreen does not have functionality for regression modelling and is impractical for very large and complex surveys where one may wish to estimate prevalence for many subsets of the data (e.g. estimating prevalence by vector species, by country/region/village, by sampling year, etc.). The R package pooling is designed primarily for case-control studies with pooled assays and not applicable to MX studies. The R packages, binGroup and its successor binGroup2, provide tools for fixed-effect regression modelling with pooled data, though only in a frequentist framework. However, neither PoolScreen nor binGroup2 have functionality to account for hierarchical sampling frames which are common in large-scale MX surveys and therefore will tend to underestimate the uncertainty in prevalence estimates associated with the sampling design. The lack of a tool that is tailored for large-scale hierarchical MX surveys limits the efficiency of data analyses and the amount of information to be gleaned from these studies, particularly in resource-poor settings with limited technical capacity.

To fill this gap, we developed PoolTestR, an package for the R language (R Core Team, 2020) which provides a user-friendly and extensible framework for estimating prevalence with pooled data, and performing fixed and mixed-effect regression modelling for both hierarchical and non-hierarchical surveys. All analyses can be conducted in frequentist or Bayesian frameworks. Though our package can be applied to generic pooled testing datasets, we demonstrate its use through examples based on simulated MX data with known prevalence.

# 2 Pooled testing and the pooled binomial GLMM

Suppose we have a molecular marker of infection in a population, with unknown prevalence $p$. We can estimate the prevalence by taking a random sample of binary outcomes showing whether or not the molecular marker is present in a sampled unit (e.g. testing each mosquito caught). If the molecular marker has a low prevalence in the population, and the unit cost of each test is much more than the unit cost of procuring samples (e.g. trapping mosquitoes), pooled testing may be a more cost-effective means of estimating prevalence. For simplicity, assume that the test can detect the marker with 100% sensitivity and specificity. We also assume that the marker is independent across each individual in the sample (and therefore also in the pools) and that the total number of samples is much smaller than the population. While larger pools may 'dilute' the marker of interest and thus affect the sensitivity of the test, for the purposes of our analysis we will assume that the dilution is insignificant for the range of pool sizes used. Under these assumptions, for a pool of size $s$, the probability of a positive result is:

$$\phi_s(p) = 1 - (1-p)^s.$$

In some cases a fixed pool size is used, however we consider the general case where there may be a mixture of pool sizes. Suppose we have a set of observations where we observe $y_i$ positive tests out of $n_i$ pools each of size $s_i$. Given the pool counts $\mathbf{n}$ and pool sizes $\mathbf{s}$, the counts of positive tests $\mathbf{y}$ is a sufficient statistic for the prevalence $p$ and follows a "pooled binomial distribution", with probability mass function given by:

$$\mathrm{PoolBin}(\mathbf{y}|\mathbf{n},\mathbf{s},p) \equiv \prod_i \mathrm{Bin}(y_i|n_i,\phi_{s_i}(p)) = (1-p)^{N-N_+} \prod_i \binom{n_i}{y_i} (1-(1-p)^{s_i})^{y_i},$$

where $N = \sum s_i n_i$ is the total number of individuals in the sample and $N_+ = \sum s_i y_i$ is the total number of individuals in all the positive pools in the sample.

Other than in trivial situations where all the positive pools have the same size, the maximum likelihood estimate (MLE) for $p$ does not have a closed-form expression and is computed numerically. Under some weak regularity conditions, the standardised MLE converges in distribution to the standard normal distribution allowing for Wald confidence intervals. Other confidence intervals for $p$ have been proposed (Hepworth, 2005).

In our package we implement a generalised linear mixed model (GLMM) using the pooled binomial distribution. Our mixed model has outcomes $\mathbf{y}_k$ that are associated with covariates $\mathbf{x}_k$ (using fixed effects) and $\mathbf{z}_k$ (using random effects).

$$f(p_k) = \mathbf{x}_k^{\mathrm{T}}\boldsymbol{\beta} + \mathbf{z}_k^{\mathrm{T}}\boldsymbol{u} \qquad \boldsymbol{u} \sim \mathrm{N}(\mathbf{0},\boldsymbol{\Sigma}),$$

where $f$ is the link function. In practice, we will generally take $\boldsymbol{\Sigma}$ to be a diagonal matrix with diagonal values (i.e., variances) that are variable parameters that may be shared over covariate groups. This yields random effect terms that are independent, but not necessarily with equal variance. Conditional on the covariates and random effects, the GLMM uses the model equation:

$$p(\boldsymbol{y}|\mathbf{n},\mathbf{s},\mathbf{x},\mathbf{z},\boldsymbol{\beta},\boldsymbol{u}) = \prod_k \mathrm{PoolBin}(\mathbf{y}_k|\mathbf{n}_k,\mathbf{s}_k,p_k).$$

This is a mixed-effects model that extends the GLM discussed in Farrington (1992); our model with a logistic link function is also similar to a bird-nesting model used in Shaffer (2004).

Our package accommodates two link functions for the prevalence parameter: the logistic and complementary log-log ($\mathrm{CLL}(\theta) = \log(-\log(1-\theta))$). The logistic link function produces more readily interpretable coefficients and is the default in our package. However, the CLL link function leads to a simpler mathematical exposition and so is illustrated here. The main benefit of the complementary log-log function in the present context is that $\mathrm{CLL}(\phi_s(\theta)) = \log(s) + \mathrm{CLL}(\theta)$, so the function can be applied to separate the pool size from the prevalence parameter. The joint log-likelihood function for the model parameters $\boldsymbol{\beta}$ and $\boldsymbol{\Sigma}$ and the random effects $\boldsymbol{u}$ is:

$$\begin{aligned}\ell_{\mathbf{y}|\mathbf{n},\mathbf{s},\mathbf{x},\mathbf{z}}(\boldsymbol{\beta},\boldsymbol{\Sigma},\boldsymbol{u}) &= \mathrm{const} + \mathrm{N}(\boldsymbol{u}|\mathbf{0},\boldsymbol{\Sigma}) \sum_k \left[(N_k - N_{k,+})\log(1-p_k) + \sum_i y_{k,i}\log\phi_{s_{k,i}}(p_k)\right] \\ &= \mathrm{const} - \mathrm{N}(\boldsymbol{u}|\mathbf{0},\boldsymbol{\Sigma}) \sum_k \Bigg[(N_k - N_{k,+})\exp(\mathbf{x}_k^{\mathrm{T}}\boldsymbol{\beta} + \mathbf{z}_k^{\mathrm{T}}\boldsymbol{u}) \\ &\qquad - \sum_i y_{k,i}\,\mathrm{log1mexp}(s_{k,i}\exp(\mathbf{x}_k^{\mathrm{T}}\boldsymbol{\beta} + \mathbf{z}_k^{\mathrm{T}}\boldsymbol{u}))\Bigg],\end{aligned}$$

where $\mathrm{log1mexp}(p) = \log(1-\exp(-p))$.

Our package implements both a frequentist and a Bayesian analysis of this GLMM. In the Bayesian case, the functions in our package have flexibility for the user to input their own prior for the unknown model parameters $\boldsymbol{\beta}$ and $\boldsymbol{\Sigma}$, or use default uninformative priors. For any prior density $\pi_0$, the posterior distribution is:

$$\pi_N(\boldsymbol{\beta},\boldsymbol{\Sigma},\boldsymbol{u}|\mathbf{y},\mathbf{n},\mathbf{s},\mathbf{x},\mathbf{z}) \propto \exp(\ell_{\mathbf{y}|\mathbf{n},\mathbf{s},\mathbf{x},\mathbf{z}}(\boldsymbol{\beta},\boldsymbol{\Sigma},\boldsymbol{u})) \cdot \pi_0(\boldsymbol{\beta},\boldsymbol{\Sigma}).$$

The covariates in these models can represent any characteristic of pools. For a MX study of a mosquito-borne disease, this may include sample collection time and location, or attributes of the site where the sample is collected (e.g. interventions in place at the site, altitude, vegetation index, distance to housing). It may also include any attributes of the mosquitoes shared by the entire pool; e.g. mosquito species, in a survey design where trapped mosquitoes are sorted by species before being pooled. However, there is no scope for units within the same pool to have different covariates unlike in BinGroup2 (Hitt et al., 2020). Mixed-effect terms can be used to account for intra-site variation not captured by other covariates in studies with hierarchical sampling frames.

## 3 The PoolTestR Package

The `PoolTestR` package has been designed to be a simple, user-friendly and extensible way to analyse test results from pooled samples. The package has four primary functions for the estimation of prevalence: `PoolPrev`, `HierPoolPrev`, `PoolReg`, and `PoolRegBayes`. `PoolPrev` produces unadjusted estimates of prevalence of a marker based on the outcome of tests on pooled samples, optionally stratifying the dataset by one or more user-specified covariates. `HierPoolPrev` is like `PoolPrev` but allows users to adjust prevalence estimates for hierarchical structure in sampling frames.

`PoolReg` and `PoolRegBayes` provide flexible and extensible frameworks to fit mixed or fixed effect regression models in either a frequentist or Bayesian framework, allowing users to identify variables associated with higher prevalence, produce predictive models, while accounting for hierarchical sampling frames. Table 1 provides an overview of the differences between the four main functions. The following sections provide more details of these functions, Boxes A and B provide example code, and Figures 1 and 2 compare the outputs of these functions when applied to a synthetic dataset.

## 3.1 PoolPrev

`PoolPrev` was designed to produce comparable results to the popular stand-alone application PoolScreen (Katholi and Unnasch, 2006) for familiarity to existing users of the software, and to enable direct comparison of results from the many studies that used PoolScreen. Stratifying a dataset and calculating prevalence for each subgroup of the data using PoolScreen requires many manual steps to import data, run analyses and export results. Using our function `PoolPrev,` this same task can be achieved in a few lines of R code with a simple syntax.

Given a dataset containing the number of samples per pool and the test results for each pool, `PoolPrev` returns Bayesian and maximum likelihood estimates of the prevalence together with uncertainty intervals. Efficient Bayesian inference is performed with Hamiltonian Markov Chain Monte Carlo using the Stan programming language (Stan Development Team, 2020b) and the R packages `rstan` (Stan Development Team, 2020a) and `rstantools` (Gabry et al., 2020). Users can specify their prior belief for the prevalence from the *Beta* distribution or use the default uninformative 'Jefferey's' prior i.e. $Beta(0.5,0.5)$. Users can also optionally specify the prior probability that the marker of interest is entirely absent from the population, in which case `PoolPrev` also returns the probability of absence given the data. As we assume the test performed on the pooled samples does not produce false positive or negatives, the probability of absence is always zero if any of the pools test positive. In most cases the credible interval (CrI) for the prevalence are the 2.5% and 97.5% quantiles of the posterior distribution. However, if all tests are positive the upper bound of the CrI is 1 and the lower bound is the 5% quantile of the posterior. Similarly, if all tests are negative the lower bound of the CrI is 0 and the lower bound is the 95% quantile of the posterior. The uncertainty interval for the maximum likelihood estimate is calculated using the likelihood ratio method (i.e. a Wilk's confidence interval). As with the Bayesian CrI, the lower or upper bound of the confidence intervals are zero or one when all pools are negative or positive, respectively.

All estimates can be optionally stratified by variables (e.g. vector species or location) by providing the name(s) of the columns in the dataset containing the variable(s). Estimation of prevalence proceed independently for each subgroup of the data defined by the variable(s).

Box A demonstrates the use of `PoolPrev` on a synthetic dataset (described in Section 4).

## 3.2 HierPoolPrev

`HeirPoolPrev` is designed to account for the hierarchical sampling structures that are common in MX studies. It assumes that samples were taken from a number of sites across the study area, and these sites can be nested within one or more hierarchical levels (e.g. sites within villages, villages within regions, regions within provinces, etc.). `HeirPoolPrev` estimates prevalence by fitting an intercept-only hierarchical generalised linear mixed model with a logistic link function. The syntax and outputs are very similar to `PoolPrev`. There is only one additional argument, `hierarchy`, which requires the user to list the variables that encode the hierarchical structure of the sampling frame (e.g. the names of columns containing site IDs, village IDs etc.). The output provides the Bayesian posterior mean and CrI for the prevalence, but unlike `PoolPrev` does not provide frequentist outputs (i.e. maximum likelihood estimates or likelihood ratio confidence intervals). As with `PoolPrev,` users can specify their prior belief for the prevalence and specify variables that stratify the dataset into subgroups. If subgroups of the data are specified, estimation of prevalence and random effect variances proceed independently for each subgroup.

Box A demonstrates the use of `HierPoolPrev` on a synthetic dataset (described in Section 4).

## 3.3 Regression - PoolReg and PoolReg Bayes

Our package provides tools for mixed-effect regression models in both frequentist and Bayesian frameworks. `PoolReg` fits a frequentist mixed- or fixed-effect generalised linear model that adjusts for the sizes of pools, building on `glm` from the in the `stats` package (R Core Team, 2020) for fixed-effect models and the `glmer` function from the `lme4` package (Bates et al., 2015) for mixed-effect models. For a model with only fixed effects the output is an S3-object of class `glmfit`, while the output for a model with random effects is an S4-object of class `glmerMod` which supports that same methods (e.g. summary, predict, plot, confint, anova) as any other object returned by the `glm` or `glmer` functions. `PoolRegBayes` provides functionality to perform the same analyses in a Bayesian framework and returns a `brmsfit` object. By building on these existing statistical packages, `PoolTestR` leverages the extensive suite of diagnostics tools available for working with models fitted with these functions and uses paradigms that will be familiar to existing users of `R`. These frameworks allow for a very broad range of linear models (e.g. polynomial regression, spline regression, gaussian process models). In addition, `PoolTestR` includes the function `getPrevalence`, which provides a convenient way to extract estimates of prevalence from regression models fitted with `PoolReg` or `PoolRegBayes`. The function `getPrevalence`, is in many cases able to detect whether a model includes adjustments for hierarchical random/group effect terms, and automatically estimate prevalence at every level of the sampling hierarchy.

Box B applies `PoolReg` and `PoolRegBayes` to the same synthetic dataset used to demonstrate `PoolPrev`, estimating the trend of decline in prevalence over time.

# 4 Comparison of methods on a synthetic dataset

`PoolTestR` provides a number of approaches to estimate prevalence: frequentist or Bayesian, stratifying or adjusting for covariates, adjusting for or ignoring hierarchical sampling frame (Table 1). We compare the approaches with a large synthetic dataset including 179,092 mosquitoes split into 7,604 pools sampled across three years with a realistic hierarchical sampling design in three regions, 30 villages (10 per region), and 300 sites (10 per village).

The synthetic dataset was generated by simulating samples taken from across three regions (A, B, and C) in which the vectors have a low (0.5%), medium (2%), and high (4%) prevalence of the marker of interest. Within each region we choose ten villages, and within each of these villages we choose ten sites to place traps. We sample from the same locations once a year over three years (0, 1, and 2). Prevalence is not uniform within each region or over time. At baseline (year 0), prevalence varies between villages within each region, and prevalence varies between sites within each village. Consequently, though the prevalence is different for each site, two sites within the same village are likely to have a more similar prevalence than two sites in different villages, or two sites in different regions. On average the prevalence is declining over time (odds ratio of 0.8 per year), however, the growth rate varies between villages. Consequently, two sites in different villages with similar prevalence at baseline may have different prevalence by the third year, and prevalence may go up in some villages.

Each year the traps at each site catch a negative binomial number (mean 200, dispersion 5) of vectors. The catch size at each site and year is independent. Though a wide range of pool sizes leads to better estimates (Gu et al., 2004), we simulate a simple and commonly used pooling strategy: each year, the catches at each site are pooled into groups of 25 with an additional pool for any remainder (e.g. a catch of 53 vectors will be pooled into two pools of 25 and one pool of three). Each pool is tested once for the marker of interest using a test with perfect sensitivity and specificity.

This synthetic dataset is provided with the PoolTestR package, and has been used to illustrate the package in Boxes A and B. Box A demonstrates the use of the functions `PoolPrev` and `HierPoolPrev`, by estimating prevalence stratified by year and region with or without adjustments for the hierarchical sampling frame. Box B demonstrates the functions `PoolReg` and `PoolRegBayes` and fits logistic-type regression models with Year and Region as covariates with and without adjustment for sampling hierarchy in frequentist and Bayesian frameworks.

The predictions for each region and year, and for a selection of villages are compared in Figures 1 and 2. Since our example has adequate sample size, the estimates using a frequentist framework are very similar to estimates in a Bayesian framework using non-informative priors, e.g. compare frequentist and Bayesian outputs of `PoolPrev` in Figures 1 and 2. For this synthetic dataset in which there is moderate variability between sites and villages, accounting for hierarchical sampling increases the fraction of confidence/credible intervals that contain the true value; the improvement can be seen whether stratifying other covariates `(Box A)`, or adjusting for them `(Box B)`. Stratifying the data by Year and Region produces estimates with wider confidence/credible intervals than in a regression framework; compare in Figures 1 and 2 `PoolPrev` to `PoolRegBayes` (without adjustment for hierarchy), or the results of `HierPoolPrev` to `PoolRegBayes` (with adjustments for hierarchy). This effect is particularly pronounced where prevalence is low (e.g. region A). Consequently, without adjustments for hierarchical sampling, using a regression framework further reduces the fraction of the confidence/credible intervals that contain the true value (Figures 1 and 2). However, the regression model with adjustments for hierarchical sampling frame (hatched square in Figures 1 and 2) has the narrowest intervals amongst models that consistently include the true prevalence values, and thus performs best on this synthetic dataset.

***Table 1*** *A summary of the four main functions in PoolTestR, with example function calls applied to a hypothetical dataset called* `Data` *with columns:* `NumInPool` *(number of specimens in each pool),* `Result` *(1/0 result of test for each pool),* `Cov1` *and* `Cov2` *(two covariate variables), and* `Level1` *and* `Level2` *(variables identifying sample location at two levels of the sampling frame hierarchy; e.g. village ID and site ID). * Applying the function* `getPrevalence` *to these outputs extracts the prevalence for each unique combination of covariates and sampling sites into a list of one or more* `data.frame` *objects.*

| Function | Example function call | Stratified or adjusted for covariates? | Adjusted for hierarchical sampling? | Bayesian / Frequentist | Output class |
|---|---|---|---|---|---|
| `PoolPrev` | `PoolPrev(Data, Result, NumInPool)` | Neither | No | Both | `tibble` |
| | `PoolPrev(Data, Result, NumInPool, Cov1, Cov2)` | Stratified | | | |
| `HierPoolPrev` | `HierPoolPrev(Data, Result, NumInPool, c('Level1','Level2'))` | Neither | Yes | Bayesian | `tibble` |
| | `HierPoolPrev(Data,Result,NumInPool,c('Level1','Level2'),Cov1,Cov2)` | Stratified | | | |
| `PoolReg` | `PoolReg(Result ~ Cov1 + Cov2, Data, NumInPool)` | Adjusted | No | Frequentist | `glmfit*` |
| | `PoolReg(Result ~ Cov1 + Cov2 + (1│Level1/Level2), Data, NumInPool)` | | Yes | | `glmerMod*` |
| `PoolRegBayes` | `PoolRegBayes(Result ~ Cov1 + Cov2, Data, NumInPool)` | Adjusted | No | Bayesian | `brmsfit*` |
| | `PoolRegBayes(Result~Cov1+Cov2+(1│Level1/Level2), Data, NumInPool)` | | Yes | | |

**Box A** Example R code applying functions PoolPrev and HierPoolPrev to a synthetic dataset, to get prevalence estimates for the whole dataset, estimates stratified by year and/or region, and estimates with and without adjustments for hierarchical sampling.

```r
#Looking at a few rows of the synthetic dataset to see structure
ExampleData[c(1:3,4001:4003),]

##      Year Region Village   Site NumInPool Result
## 1       0      A     A-1  A-1-1        25      0
## 2       0      A     A-1  A-1-1        25      0
## 3       0      A     A-1  A-1-1        25      0
## 4001    0      B     B-6 B-6-10         5      0
## 4002    1      B     B-6 B-6-10        25      1
## 4003    1      B     B-6 B-6-10        25      0

#Prevalence across the whole synthetic dataset (ignoring hierarchical sampling)
Prevs <- PoolPrev(ExampleData, Result, NumInPool)

#Prevalence for each Region (ignoring hierarchical sampling)
PrevsRegion <- PoolPrev(ExampleData, Result, NumInPool, Region)

#Prevalence for each Year (ignoring hierarchical sampling)
PrevsYear <- PoolPrev(ExampleData, Result, NumInPool, Year)

#Prevalence for each combination of Region and Year (ignoring hierarchical sampling)
PrevsYearRegion <- PoolPrev(ExampleData, Result, NumInPool, Region, Year)
PrevsYearRegion

## # A tibble: 9 x 11
##   Region  Year PrevMLE   CILow  CIHigh PrevBayes   CrILow CrIHigh ProbAbsent
## 1 A          0 0.00617 0.00511 0.00737   0.00621 0.00508 0.00740 NA
## 2 A          1 0.00461 0.00368 0.00569   0.00463 0.00364 0.00569 NA
## 3 A          2 0.00595 0.00492 0.00711   0.00598 0.00493 0.00710 NA
## 4 B          0 0.0197  0.0176  0.0220    0.0198  0.0176  0.0221  NA
## 5 B          1 0.0196  0.0175  0.0218    0.0196  0.0177  0.0217  NA
## 6 B          2 0.0172  0.0152  0.0194    0.0172  0.0153  0.0194  NA
## 7 C          0 0.0380  0.0348  0.0414    0.0380  0.0348  0.0414  NA
## 8 C          1 0.0303  0.0274  0.0334    0.0304  0.0277  0.0335  NA
## 9 C          2 0.0260  0.0235  0.0286    0.0260  0.0235  0.0286  NA
## # … with 2 more variables: NumberOfPools <int>, NumberPositive <dbl>

#Prevalence for each Region and Year accounting for hierarchical sampling
HierPrevsYearRegion <- HierPoolPrev(ExampleData, Result, NumInPool,
                                    c("Village", "Site"), Region, Year)
#Similar to the above but stratifying down to village level
HierPrevsYearRegionVillage <- HierPoolPrev(ExampleData, Result, NumInPool,
                                           c("Site"), Region, Year, Village)
```

**Box B** Mixed and fixed effect regression modelling for pooled data using PoolReg and PoolRegBayes

```
# Logistic regression model - no adjustment for sampling frame hierarchy
ModFreq <- PoolReg(Result ~ Region + Year,
                   ExampleData, NumInPool)
coefficients(ModFreq) # estimated model coefficients
## (Intercept)     RegionB     RegionC        Year
##  -5.0555273   1.2260503   1.7476965  -0.1287181

# Logistic regression model - adjusting for sampling frame hierarchy
ModFreqHier <- PoolReg(Result ~ Region + Year + (1|Village/Site),
                       ExampleData, NumInPool)

# Same as above but in a Bayesian framework
ModBayes <- PoolRegBayes(Result ~ Region + Year,
                         ExampleData, NumInPool)
ModBayesHier <- PoolRegBayes(Result ~ Region + Year + (1|Village/Site),
                             ExampleData, NumInPool)

# A more complex model: estimate temporal trend for each village
ModBayesHier2 <- PoolRegBayes(Result ~ Region + Year + (1 + Year|Village) + (1|Site),
                              ExampleData, NumInPool)

# We can use any of these models to estimate prevalence e.g.
getPrevalence(ModFreq)
## $PopulationEffects
##   Region Year    Estimate       CILow      CIHigh
## 1      A    0 0.006333634 0.005617496 0.007140413
## 2      A    1 0.005572933 0.004985287 0.006229414
## 3      A    2 0.004903144 0.004335404 0.005544817
## 4      B    0 0.021259202 0.019664559 0.022980127
## omitted 5 more rows

# For hierarchical models, getPrevalence returns prevalence at every level
getPrevalence(ModBayesHier2)
## $PopulationEffects
##   Region Year    Estimate      CrILow     CrIHigh
## 1      A    0 0.005233794 0.003579397 0.007534288
## 2      A    1 0.004482318 0.003065639 0.006399554
## 3      A    2 0.003842876 0.002572213 0.005569832
## 4      B    0 0.019507027 0.013687156 0.026432815
## omitted 5 more rows
## $Village
##    Region Year Village    Estimate      CrILow     CrIHigh
## 1       A    0     A-1 0.003073011 0.001690074 0.004999869
## 2       A    1     A-1 0.002665134 0.001510287 0.004283468
## 3       A    2     A-1 0.002342421 0.001232661 0.003917210
## 4       A    0     A-2 0.011871139 0.007584781 0.017769973
## omitted 86 more rows
## $Site
##     Region Year Village    Site    Estimate       CrILow     CrIHigh
## 1        A    0     A-1   A-1-1 0.002662369 0.0008586564 0.005724619
## 2        A    1     A-1   A-1-1 0.002310121 0.0007825112 0.005028522
## 3        A    2     A-1   A-1-1 0.002031356 0.0006330185 0.004518543
## 4        A    0     A-1   A-1-2 0.002877637 0.0010513109 0.005875640
## omitted 896 more rows

# We can also predict prevalence for new datapoints. For example, this projects
# the temporal trend forward to estimate prevalence in region A for years 3-5.
PredictData <- data.frame(Region = "A", Year = c(3,4,5))
getPrevalence(ModFreq, newdata = PredictData)
## $PopulationEffects
##   Region Year    Estimate       CILow      CIHigh
## 1      A    3 0.004313506 0.003711471 0.005012704
## 2      A    4 0.003794505 0.003148847 0.004571946
## 3      A    5 0.003337741 0.002658735 0.004189427
```

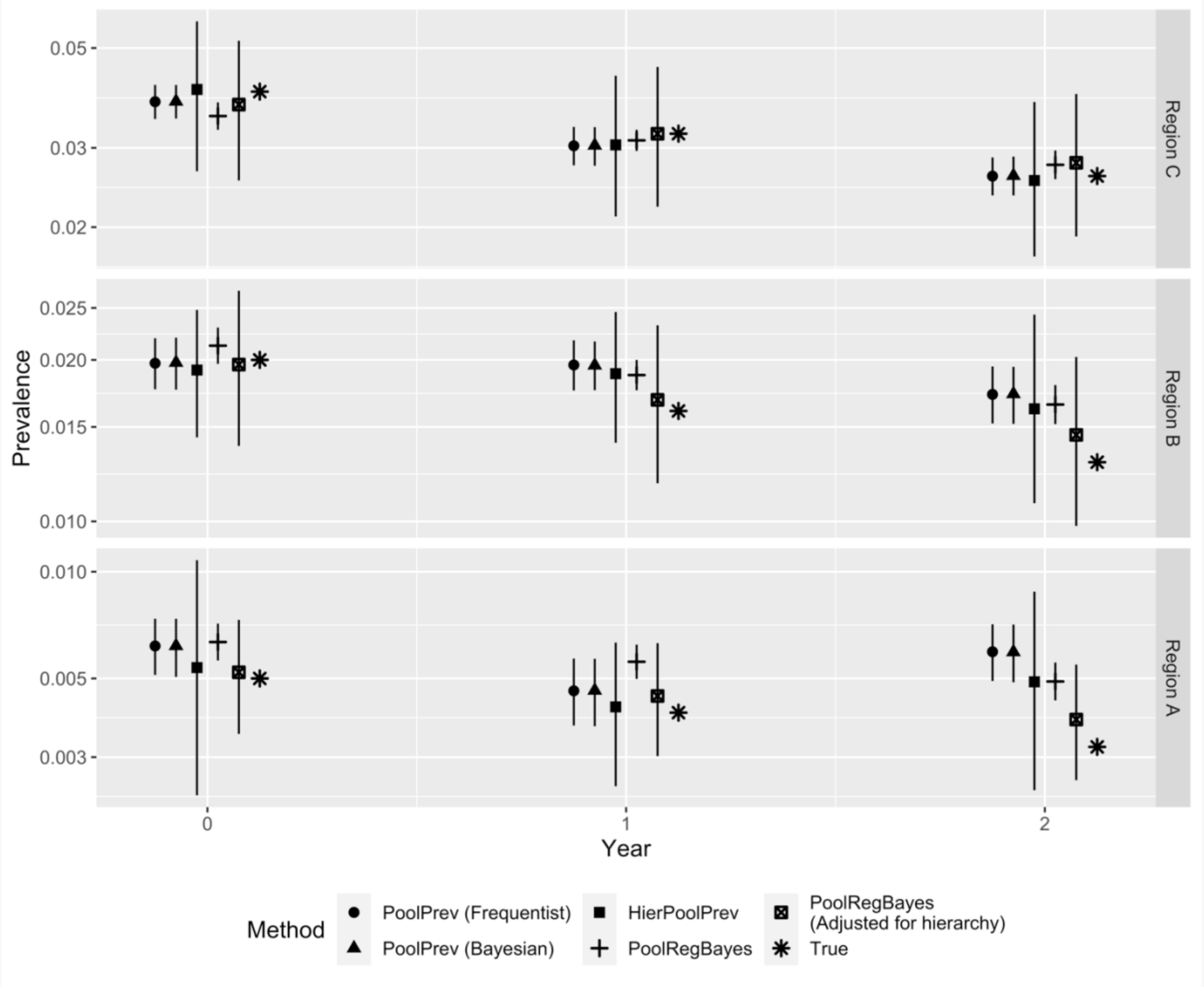

***Figure 1*** *Comparison of five different methods for estimating prevalence of a maker by region and year using a synthetic dataset of pooled samples collected over three years from across 300 sites within 30 villages within 3 regions. The vertical lines indicate 95% confidence/credible intervals. The true prevalence for each region and year is indicated by an asterisk. Only methods that account for the hierarchical nature of the sampling frame – using functions HierPoolPrev (solid square) or PoolRegBayes with random/group effect terms (hatched square) – consistently include the true value within the estimate intervals. R code for generating all estimates can be found in Boxes A and B. The function PoolPrev (circle and triangle) uses the same underlying methodology as the software PoolScreen (Katholi and Unnasch, 2006).*

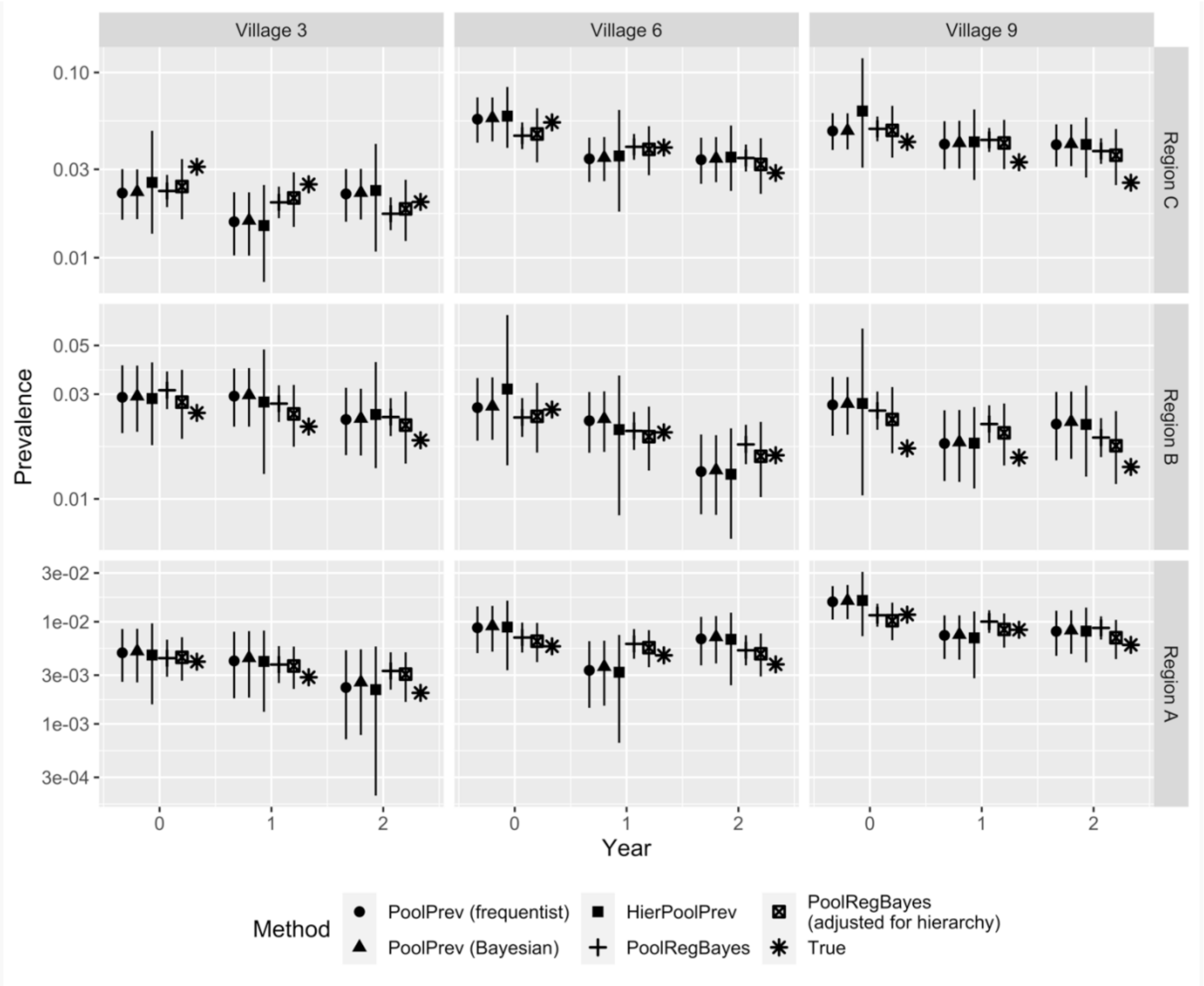

***Figure 2*** *Comparison of five different methods for estimating prevalence of a maker by village and year using a synthetic dataset of pooled samples collected over three years from across 300 sites within 30 villages within 3 regions. Only results for three villages in each region are displayed for illustrative purposes. The vertical lines indicate 95% confidence/credible intervals. The true prevalence for each region and year is indicated by an asterisk. Only methods that account for the hierarchical nature of the sampling frame – using functions HierPoolPrev (solid square) or PoolRegBayes with random/group effect terms (hatched square) – consistently include the true value within the estimate intervals. The function PoolPrev (circle and triangle) uses the same underlying methodology as the software PoolScreen (Katholi and Unnasch, 2006).*

# 5 Discussion

PoolTestR is a user-friendly, flexible, and extensible R package for estimating prevalence and regression modelling with tests on pooled samples. PoolTestR offers substantial advantages over existing software for pooled testing such as Poolscreen (Katholi and Unnasch, 2006), especially for sampling design such as those used in MX surveys. While each analysis in PoolScreen requires many manual steps to import data and export results, analysis with PoolTestR users can utilise the diverse ecosystem of R packages to easily import data, perform multiple analyses, and export results with a number of common formats (e.g. csv, xls). PoolTestR also expands the range of statistical analyses that can be performed, allowing estimates to be adjusted for hierarchical sampling design and providing tools for a very broad category of mixed effect regression models in Bayesian or frequentist frameworks.

When conducting MX surveys, collecting a simple random sample of vectors across a large area is operationally infeasible. Many MX studies will therefore involve a hierarchical sampling frame involving representative sample sites distributed across the study area. If the study area and the distance between traps are smaller than the movement range of the vector being studied, it may be fair to assume that all traps are sampling from the same population, and that there is no variation in prevalence between trap sites. In such cases the method implemented in Poolscreen and the PoolPrev function in our package are appropriate for estimating prevalence. However, when aggregating data to estimate prevalence in a study area substantially larger than the typical movement range of vectors, these methods which do not account for heterogeneity between sample sites may have unreasonably narrow confidence intervals that often fail to contain the true value (Birkner et al., 2013). Instead, the function HierPoolPrev or a hierarchical mixed-effect regression model using PoolReg or PoolRegBayes should be preferred in these situations. While accounting for hierarchical sampling frames will increase the width of confidence intervals for prevalence estimates, failing to do so may result in confidence intervals which frequently fail to include the true prevalence value.

Molecular xenomonitoring surveys utilising pooled testing are often paired with human surveys utilising un-pooled testing. Though regression modelling is commonly used in the human components of these surveys (e.g. Subramanian et al. (2020)), regression modelling with pooled MX data has been hampered by the lack of suitable software; the only method for looking at differences by groups in PoolScreen is to stratify the data, and the regression models in binGroup2 cannot account for hierarchical sampling frames. The regression functions in the PoolTestR package fill this gap, allowing users to identify variables associated with higher prevalence, test the statistical significance of these associations, and produce predictive models. Moreover, where appropriate regression models can produce more precise estimates (narrower confidence intervals) compared to simple stratification. Regression models could be used for predictive prevalence mapping, however further development is required to allow for models with spatial correlation in our package. There are currently no tools that readily allow for the comparison or synthesis of both the human and MX components of surveys (e.g. model predictions of prevalence in humans based on prevalence in vectors). This functionality may be added in future releases of PoolTestR.

As with all models, estimates made with PoolTestR will be unreliable if the implicit assumptions about the test characteristics, sampling frame, population, or covariates are substantially violated. All the models in our package currently assume that the tests applied to each pool has perfect sensitivity and specificity. One of many things that can affect test sensitivity and specificity, and therefore the accuracy of the prevalence estimates, is pool size. Statistical methods that estimate test sensitivity or specificity from the data or test for the existence of diluting effect in larger pools have been proposed (Tu et al., 1995) and may be incorporated in future versions of PoolTestR. All of our models also assume that vector catch numbers are fixed by the sampling design or that the 'stopping rules' used to determine samples sizes are independent of prevalence and covariates. A common 'stopping rule' is to test as many vectors as possible from each site. The relationship between vector density, transmission rate, and prevalence is dependent on complex host, agent, and environment relationships, and so there may be correlation between catch numbers and disease prevalence at a site. This kind of correlation, if not accounted for, will result in biased estimates, if testing all vectors trapped at each site. While a predetermined sample size for each site could avoid this bias, it may require sampling to be prolonged at some sites and vectors to be discarded at others. The best way to detect and adjust for bias related to sampling designs that do not use a predetermined sample size remains an open question.

Another key consideration in MX studies is the appropriate sample size and pooling strategy. When designing a sampling strategy using pooled samples, there is a trade-off between cost and precision. Using fewer, larger pools makes it cheaper and faster to conduct laboratory tests, but greater numbers of smaller pools improves the power of the data and the precision of estimates. For a fixed number of pools, distributing the specimens into a number of fixed size pools is likely to result in poorer estimates than using pools of various sizes (Gu et al., 2004). However, there are currently no practical rules or tools for determining an optimal or near-optimal strategies for sampling or pooling. A tool that — given a sampling design, testing constraints, and catch size — determines the optimal number of pools and the optimal distribution of samples across these pools, would further improve the cost-effectiveness of pooled MX surveys and may be incorporated in future updates of PoolTestR.

# 6 Conclusion

PoolTestR is a software package borne out of the need for a simple, flexible and freely available tool to analyse large and complex datasets to estimate infection prevalence from pooled samples. PoolTestR allows users to conduct the most common analyses required for MX, whilst to being able to adjust for hierarchical sampling design and conduct a broad range of regression analyses. MX is increasingly being used as a surveillance method around the world and we hope that PoolTestR can assist researchers and program managers in disease surveillance in a range of control settings and others contexts using pooled data.

# 7 Acknowledgements:

Angus McLure was supported by an Australian Research Council Discovery Project Grant (DP180100246). Colleen Lau was supported by an Australian National Health and Medical Research Council Fellowship (1109035).

This work received financial support from the Coalition for Operational Research on Neglected Tropical Diseases (COR-NTD) (Grant number OPP1053230), which is funded at The Task Force for Global Health primarily by the Bill & Melinda Gates Foundation, by the UK aid from the British government, and by the United States Agency for International Development through its Neglected Tropical Diseases Program.

The funders had no role in study design, data collection and analysis, decision to publish, or preparation of the manuscript.